\documentclass{elsart}

\usepackage{natbib}

\usepackage{amssymb}
\usepackage{graphicx}
\newcommand{\lora} {\boldmath$\longrightarrow$}
\newcommand{\vecbm}[1]{\mbox{\boldmath#1}}

\newcommand{\cent}[1] {\begin{center}#1\end{center}}

\begin{document}

\begin{frontmatter}



\title{Classical Equilibrium Thermostatistics,\\ "Sancta
sanctorum of Statistical Mechanics\protect\footnote{as coined by C.Tsallis}"\\
From Nuclei to Stars}

\author{D.H.E. Gross}

\address{Hahn-Meitner Institut and Freie Universit{\"a}t Berlin,
Fachbereich Physik, Glienickerstr. 100, 14109 Berlin, Germany, and
Universit\`a di Firenze and INFN, Sezione di Firenze, via Sansone 1, 50019
Sesto F.no (Firenze), Italy.}

\begin{abstract}
% Text of abstract
Equilibrium statistics of Hamiltonian systems is correctly described
  by the microcanonical ensemble. Classically this is the manifold of all
  points in the $N-$body phase space with the given total energy.
  Due to Boltzmann-Planck's principle, $e^S=tr(\delta(E-H))$, its
  geometrical size is related to the entropy $S(E,N,V,\cdots)$. This
  definition does not invoke any information theory, no
  thermodynamic limit, no extensivity, and no homogeneity
  assumption.  Therefore, it describes the equilibrium statistics of extensive
  as well of non-extensive systems. Due to this fact it is {\em the
  fundamental} definition of any classical equilibrium statistics. It
  addresses nuclei and astrophysical objects as well. $S(E,N,V,\cdots)$ is
  multiply differentiable everywhere, even at phase-transitions.  All kind of
  phase transitions can be distinguished sharply and uniquely for even small
  systems. What is even more important, in contrast to the canonical theory,
  also the region of phase-space which  corresponds to phase-separation is
  accessible, where the most interesting phenomena occur. No deformed
  q-entropy is needed for equilibrium. Boltzmann-Planck is the only
  appropriate statistics independent of whether the system is small or large,
  whether the system is ruled by short or long range forces.
\end{abstract}


\end{frontmatter}

% main text
\section{Introduction}
\vspace*{-0.5cm}
Classical Thermodynamics was originally designed to describe
steam engines. It got its theoretical foundation in classical thermostatistics.
Now Thermodynamics and the theory of phase transitions of
homogeneous and large systems are some of the oldest and best
established theories in physics. It may look surprising to add
anything new to it. Let me recapitulate what was told us since
$\sim 150$ years:
\begin{itemize}
\item Thermodynamics addresses large homogeneous systems at equilibrium
(in the thermodynamic limit
$N\to\infty|_{ N/V=\rho, homogeneous}$).
\item Phase transitions are the positive zeros of the grand-canonical
partition sum $Z(T,\mu,V)$ as function of $e^{\beta\mu}$
(Yang-Lee-singularities). Of course these singularities indicate the break-down
of the (grand-)canonical ensemble.
\item Micro and canonical ensembles are equivalent.
\
\item Thermodynamics works with intensive variables $T,P,\mu$.
\item Unique Legendre mapping $T\to E$.
\item Heat only flows from hot to cold (Clausius).
\item Second Law only in infinite systems when the Poincar\'{e}
recurrence time becomes infinite (much larger than the age of the
universe (Boltzmann)).
\end{itemize}

 Under these constraint only a tiny part of the real world of
equilibrium systems can be treated. The ubiquitous non-homogeneous
systems: nuclei, clusters, polymers, soft matter (biological)
systems, but also the largest, astrophysical systems are not
covered. Even normal systems with short-range coupling at phase
separations are inhomogeneous and can only be treated within
conventional homogeneous thermodynamics (e.g. in van-der-Waals
theory) by bridging the unstable region of negative
compressibility by a Maxwell construction. Thus even
the original goal, for which Thermodynamics was invented some
$150$ years ago, the description of steam engines is only
artificially solved. There is no (grand-)canonical ensemble of
phase separated and, consequently, inhomogeneous, configurations.
This has a deep reason as I will discuss below: here the systems
have a {\em negative} heat capacity $c$ (resp. susceptibility). This,
however, is impossible in the (grand-)canonical theory where
$c\propto (\delta E)^2$

As was recently remarked by Pitowsky\cite{pitowsky01}: "There is a
schizophrenic attitude in the foundations of statistical mechanics. While
Boltzmann's view has been promoted as conceptional superior to that of
Gibbs, it is the canonical and not the microcanonical probability
distribution that is extensively used in calculations. The claim being that
the mathematical simplifications aside, all the ensembles are
thermodynamically equivalent because as long we are dealing in systems
containing a large number of molecules.". Here, I will show that this is
not only schizophrenic but is merely wrong in the most interesting
situations. And that not only for small systems but also for the really
large ones.
\section{Boltzmann-Planck's principle}
\vspace*{-0.5cm}
   The Microcanonical ensemble is the ensemble (manifold)
   of all possible points in the $6N$ dimensional phase space at
   the prescribed sharp energy $E$:
\begin{eqnarray*}
W(E,N,V)&=&\epsilon_0 tr\delta(E-H_N)=\epsilon_0\int{\frac{d^{3N}p\;d^{3N}q}{N!(2\pi\hbar)^{3N}}
\delta(E-H_N)}.
\end{eqnarray*}
Thermodynamics addresses the whole ensemble. It is ruled by
  the topology of the geometrical size $W(E,N,\cdots)$, as
  expressed on Boltzmann's epitaph:
 \begin{equation}
\fbox{\fbox{\vecbm{S=k*lnW}}}
\end{equation}
which is the most fundamental definition of the entropy $S$.
Entropy and with it micro-canonical thermodynamics has a pure
mechanical, geometrical foundation. No information theoretical
formulation is needed. Moreover, in contrast to the canonical
entropy, $S(E,N,..)$ is everywhere single valued and multiple
differentiable. There is no need for extensivity, for
concavity, for additivity, and no need for the thermodynamic
limit. This is a great advantage of the geometric foundation of
equilibrium statistics over the conventional definition of the
Boltzmann-Gibbs (BG) canonical theory. However, addressing entropy to
finite eventually small systems we will face a new problem with
Zermelo's objection against the monotonic rise of entropy, the
Second Law. Here the Poincar\'{e} recurrence time might be small and
Boltzmann's excuse does not work anymore. This is
discussed elsewhere \cite{gross183,gross192,gross174}.

\section{Topological properties of $S(E,\cdots)$}
\begin{figure}[b]
\includegraphics*[bb =0 0 290 180, angle=0, width=14cm,
clip=true]{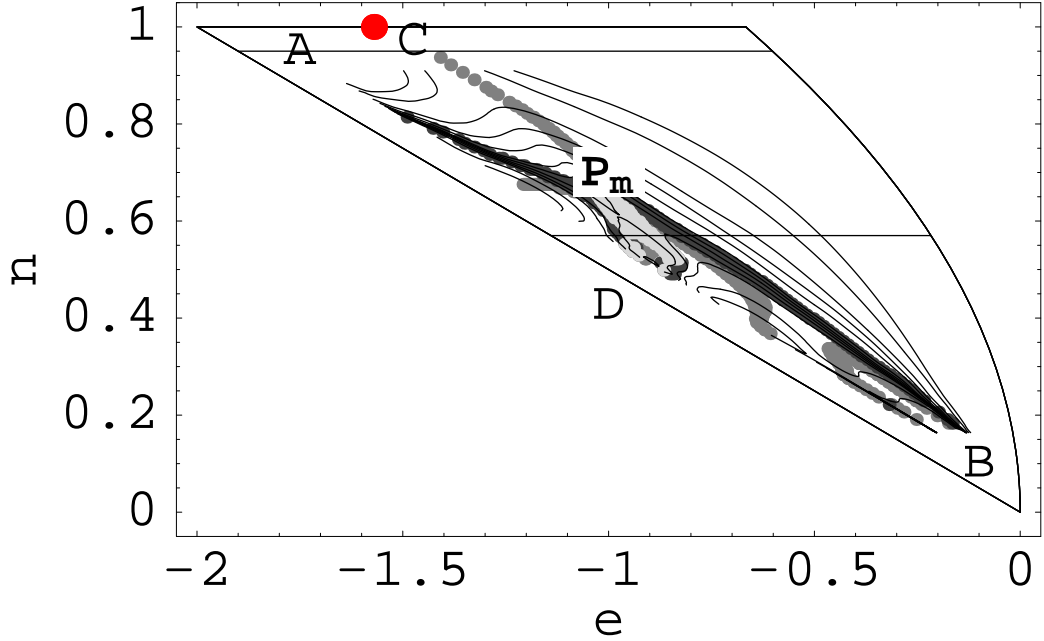}
\caption{{\bf Global phase diagram or contour plot of the curvature
determinant (Hessian) for a small system}, eqn.~(\ref{curvdet}),
of the 2-dim Potts-3 lattice gas
  with $50*50$ lattice points, $n$ is the number of particles per
  lattice point, $e$ is the total energy per lattice point. The line
   (-2,1) to (0,0) is the ground-state energy of the lattice-gas
   as function of $n$. The most right curve is the locus of configurations
    with completely random spin-orientations (maximum entropy). The whole
    physics of the model plays between these two boundaries. The different
  regions of positive or negative Hessian, i.e. of respectively negative or positive
  maximum curvature $\lambda_1$ correspond to different phases (in the case
  studied here $\lambda_2$ is everywhere negative):  At the
    dark-gray (green in the color version) lines the Hessian is $\det=0$,
    this is the boundary of the region of phase separation (the triangle
    $AP_mB$) with a negative Hessian ($\lambda_1>0,\lambda_2<0$).  Here, we have
    Pseudo-Riemannian geometry and a convex entropy. At the light-gray
    (blue in the color version)
    lines is a minimum of $\det(e,n)$ in the direction of the largest
    curvature (\vecbm{v}$_{\lambda_1}\cdot$\vecbm{$\nabla$}$\det=0$) and additionally
    $\det\approx 0$, these are lines of second order transition. In the triangle
    $AP_mC$ is the pure ordered (solid) phase with concave entropy and
    ($\det>0, \lambda_1<0,\lambda_2<0$). Above and right of the line $CP_mB$ is the
    pure disordered (gas) phase again with concave $s(e,n)$ and
    ($\det>0, \lambda_1<0,\lambda_2<0$). The crossing $P_m$ of the boundary lines is a
    multi-critical point. Here we have simultaneously
    $\lambda_1=0,$\vecbm{$\nabla$}$\lambda_1=0$. It is also the critical end-point
    of the region of phase separation.  The lighter-gray (red in the color version)
    region around the multi-critical point $P_m$ corresponds to a flat,
    horizontal region of $\det(e,n),\lambda_1\sim 0$  and consequently
    to a somewhat extended cylindrical region of $s(e,n)$,
    see \protect\cite{gross174,gross173}; $C$ (red) is the analytically
   known position of the critical point (second order transition) which the
   ordinary $q=3$ Potts model (without vacancies) {\em would have in the
   thermodynamic limit}. $C$ is also approached by the line of second-order
   transition (light-gray/blue) in the present small system.}\label{det}
   %%%and {\boldmath\vecbm{$\nabla$}\mbox{$\lambda_1$}{\boldmath$\sim 0$}}
\end{figure}

\vspace*{-0.5cm}
The topology of the Hessian of $S(E,\cdots)$, the determinant of
curvatures of $S(E,\cdots)$, determines completely all kinds of phase
transitions. This is evidently so, because $e^{S(E)-E/T}$ is the
weight of each energy in the canonical partition sum at given $T$.
Consequently, at phase separation this
has at least two maxima, the two phases. And in between two maxima
there must be a minimum where the curvature of $S(E)$ is positive.
I.e. the positive curvature detects phase separation. This is of
course true also in the case of several conserved control parameters
e.g. $e=E/L^2,n=N/L^2,s=S/L^2$,  in the case of the $q=3$
Potts-gas model on a two dimensional lattice of {\em finite} size
of $L^2=50\times 50$ lattice points. Here the Hessian is [$\lambda_i$
are the eigenvalues = eigen-curvatures]:
\begin{eqnarray}
d(e,n)&=&\left\|\begin{array}{cc} \frac{\partial^2 s}{\partial
e^2}& \frac{\partial^2 s}{\partial n\partial e}\\ \frac{\partial^2
s}{\partial e\partial n}& \frac{\partial^2 s}{\partial n^2}
\end{array}\right\|=\lambda_1\lambda_2 \label{curvdet}\\
\lambda_1&\ge&\lambda_2\hspace{1cm}\mbox{\lora eigenvectors
:}\hspace{1cm} {\boldmath\vecbm{$v$}_1,\vecbm{$v$}_2}\nonumber
\end{eqnarray}
The whole zoo of phase-transitions can be sharply seen and distinguished
[fig.(\ref{det})].
\subsection{Curvature}
\begin{figure}[h]
\includegraphics*[bb = 99 57 400 286, angle=-0, width=9cm,
clip=true]{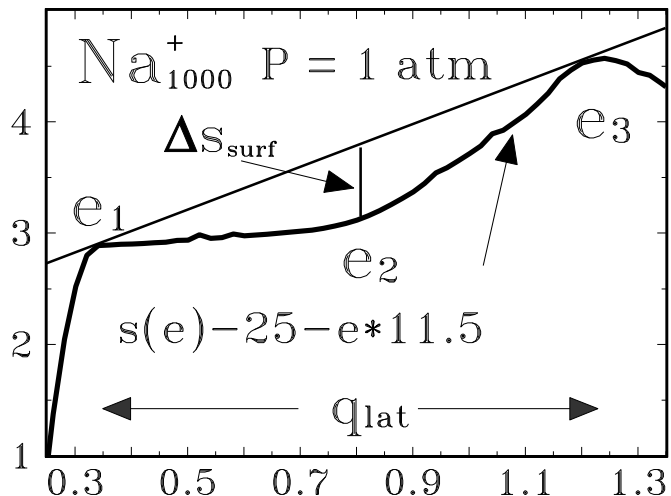}
\caption{{\bf The physical origin of convex regions in $S(E)$ for systems with
interactions of short range.} MMMC~\protect\cite{gross174}
simulation of the entropy $s(e)$ per atom ($e$ in eV per atom) of
a system of
  $N_0=1000$ sodium atoms at an external pressure of 1 atm.  At the
  energy $e\leq e_1$ the system is in the pure liquid phase and at
  $e\geq e_3$ in the pure gas phase, of course with fluctuations which are
  proportional to the inverse negative curvature of $s(e)$. The
  latent heat per atom is $q_{lat}=e_3-e_1$.  \underline{Attention:}
  the curve $s(e)$ is artifically sheared by subtracting a linear
  function $25+e*11.5$ in order to make the convex intruder visible.
  {\em $s(e)$ is always a steeply monotonic rising function}.  We
  clearly see the global concave (downwards bending) nature of $s(e)$
  and its convex intruder. Its depth is the relative entropy
  loss due to additional correlations by internal interfaces. It scales $\propto
  N^{-1/3}$. From this one can calculate the surface tension per surface atom
  $\sigma_{surf}/T_{tr}=\Delta s_{surf}*N_0/N_{surf}$. For the present example
  it is $\sigma_{surf}/T_{tr}=5.68$ which should be compared to
  $\sigma_{surf}/T_{tr}=9.267$ for the bulk. The bulk value is approximated by
  $\sigma_{surf}/T_{tr}$ systematically with rising number of atoms $N$
  [more details c.f.\cite{gross174,gross189}].  The double
  tangent (Gibbs construction) is the concave hull of $s(e)$. Its
  derivative gives the Maxwell line in the caloric curve $T(e)$ at
  $T_{tr}$. In the thermodynamic limit the intruder would disappear and $s(e)$
  would approach the double tangent from below.  Nevertheless, even
  there, the probability $\propto e^{Ns}$ of configurations with
  phase-separations are suppressed by the
  (infinitesimal small) factor $e^{-N^{2/3}}$ relative to the pure
  phases and the distribution remains {\em strictly bimodal in the
   canonical ensemble}. The region $e_1<e<e_3$ of phase separation
 gets lost.\label{naprl0f}}
\end{figure}

\begin{figure}[h]\hspace*{-1cm}
%%%\begin{center}
\begin{minipage}{8.5cm}
\includegraphics*[bb =48 48 387 611, angle=-180, width=8.5 cm,
clip=true]{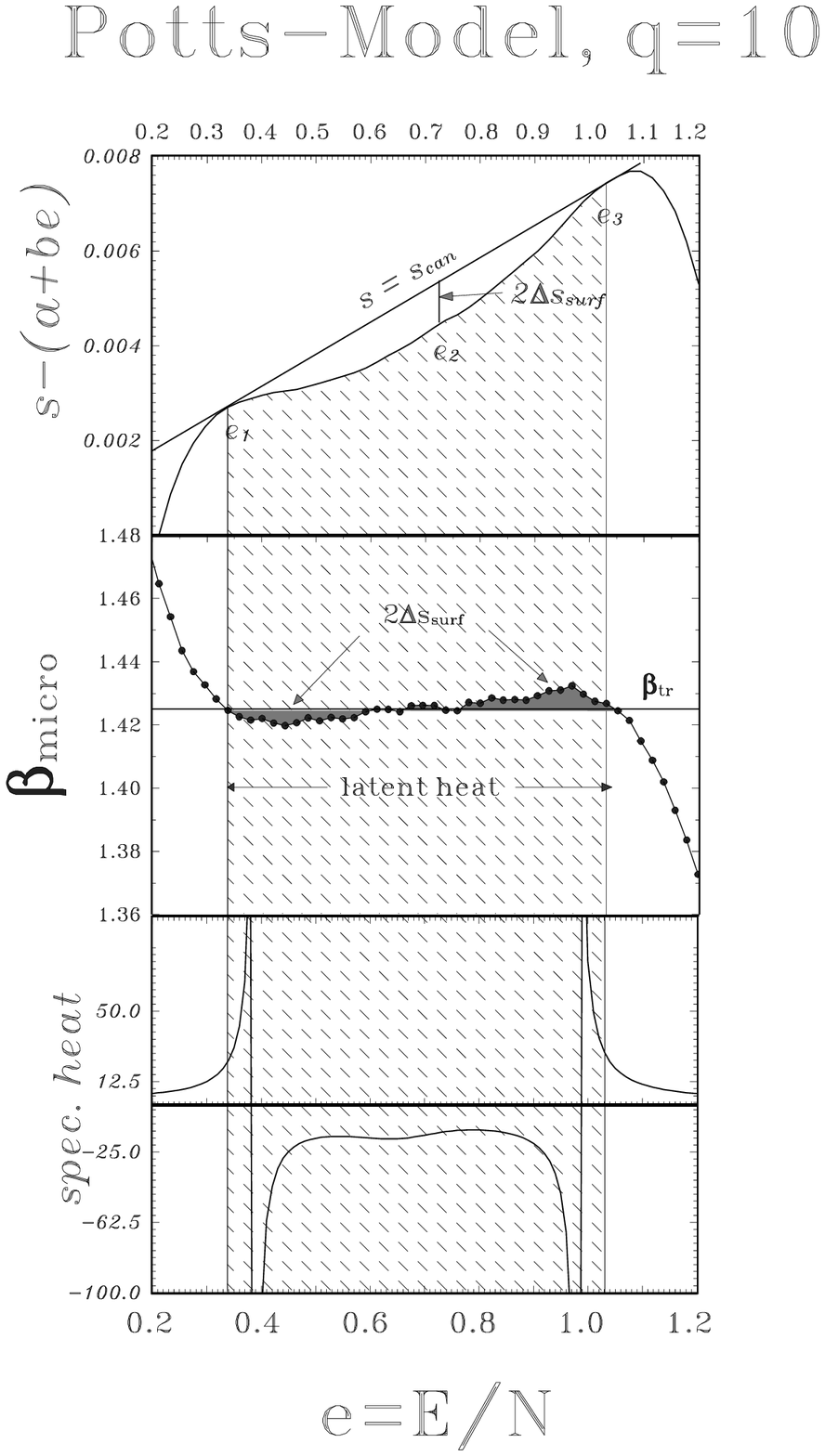}%%38 8 387 611

\end{minipage}~~~\begin{minipage}{6cm}\vspace*{-2cm}\caption{{\bf Heat can flow
from hot to cold.} Potts model (here with $q=10$ showing a strong
first order transition), in the region
of phase separation. At $e_1$ the system is in the pure ordered
phase, at $e_3$ in the pure disordered phase. A little above $e_1$
the temperature $T=1/\beta$ is higher than a little below $e_3$.
Combining two parts of the system: one at the energy $e_1+\delta
e$ and at the temperature $T_1$, the other at the energy
$e_3-\delta e$ and temperature $T_3<T_1$. It will equilibrize
with a rise of its entropy, under a dropping of $T_1$ (cooling) and an
energy flow (heat) from $3\to 1$: i.e.: Heat flows from cold to
hot! Clausius formulation of the Second Law is violated.
Evidently, this is not any peculiarity of gravitating systems!
This is a generic situation within classical thermodynamics
even for short-range coupling. It has nothing to do with
long range interactions}.\label{heat}
\end{minipage}%%%%%%\end{center}
\end{figure}

\vspace*{-0.5cm}
The curvature (Hessian) of $S(E,N,\cdots)$ controls
the phase transitions see ref.\cite{gross174}. What is the physics
behind a positive curvature?
For a short-range force it is linked to the interphase surface
tension c.f. in fig.\ref{naprl0f} and to a {\em negative heat capacity}.
This implys that {\em heat can flow from cold to hot}, c.f. fig\ref{heat}.

\subsection{Atomic clusters}
\begin{figure}[h]
 \cent{\includegraphics [bb = 69 382 545
766, angle=-0, width=12cm, clip=true]{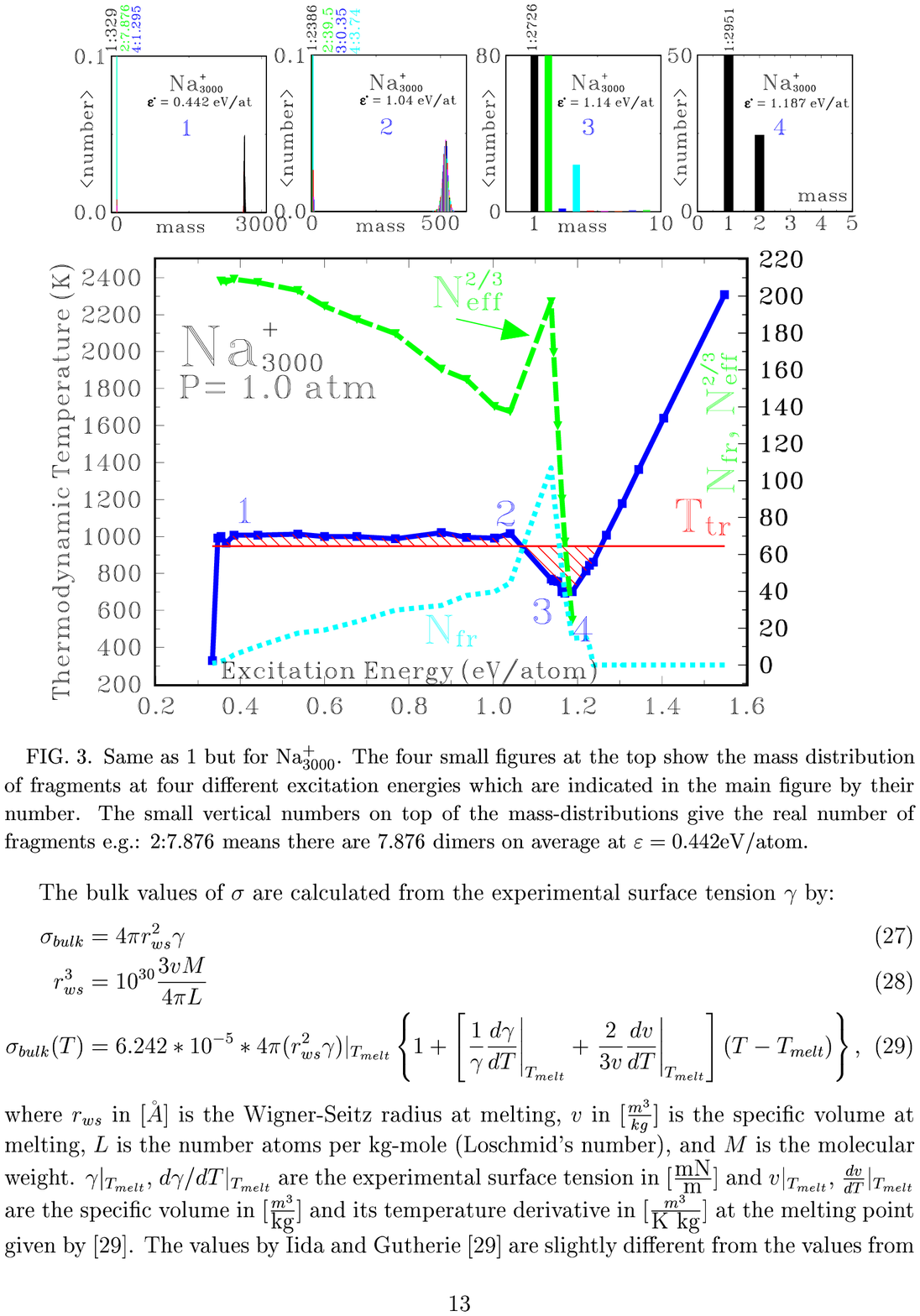}}
\caption{{\bf Atomic cluster fragmentation in the convex region
of $S(E)$.} The backbending caloric curve $T(E)$ (blue in the color version),
left scale. The number of fragments $N_{fr}$ and the number of surface atoms
$N_{eff}^{2/3}=\sum_{N_i\ge 2}N_i^{2/3}$ of larger fragments (green in the
color version) can
be read off from the right scale. $T_{tr}$ is the "Maxwell" line
dividing $T(E)$ into two equal area parts. The inserts
above give the mass distribution at the various points along the caloric
curve $T(E)$. The label
"4:1.295" means 1.295 quadrimers on average.  This gives a
detailed insight into what happens with rising excitation energy
over the transition (convex) region: At the beginning ($e^*\sim 0.442$ eV)
the liquid sodium drop evaporates 329 single atoms and 7.876
dimers and 1.295 quadrimers on average. At energies $e>\sim 1$eV
the drop starts to fragment into several small droplets
("intermediate mass fragments") e.g. at point 3: 2726 monomers, 80
dimers, $\sim$5 trimers, $\sim$15 quadrimers and a few heavier ones
up to 10-mers. The evaporation residue disappears. This
multifragmentation finishes at point 4. It induces the strong
backward swing of the caloric curve $T(E)$. Above point 4 one has
a gas of free monomers and at the beginning a few dimers. This
transition scenario has a lot similarity with nuclear
multifragmentation. It is also shown how the total interphase
surface area, proportional to $N_{eff}^{2/3}=\sum_i N_i^{2/3}$
with $N_i\ge 2$ ($N_i$ the number of atoms in the $i$th cluster)
stays almost constant up to point 3 even though the number of
fragments ($N_{fr}=\sum_i$) is monotonic rising with increasing
excitation.\label{clusterfragmentation}}
\end{figure}

\vspace*{-0.5cm}
In fig.\ref{clusterfragmentation} I show the simulation of a typical
fragmentation transition
of a system of $3000$ sodium atoms interacting by realistic
(many-body) forces. To compare with usual macroscopic conditions,
the calculations were done at each energy using a volume $V(E)$
such that the microcanonical pressure $P=\frac{\partial
S}{\partial V}/\frac{\partial S}{\partial E}=1$atm. Evidently,
the convex region of $S(E,P)$ is the most interesting region.

%%\clearpage

\subsection{Stars}
\begin{figure}[h]
    \includegraphics*[bb =88 401 522 630, angle=-0, width=12 cm,
    clip=true]{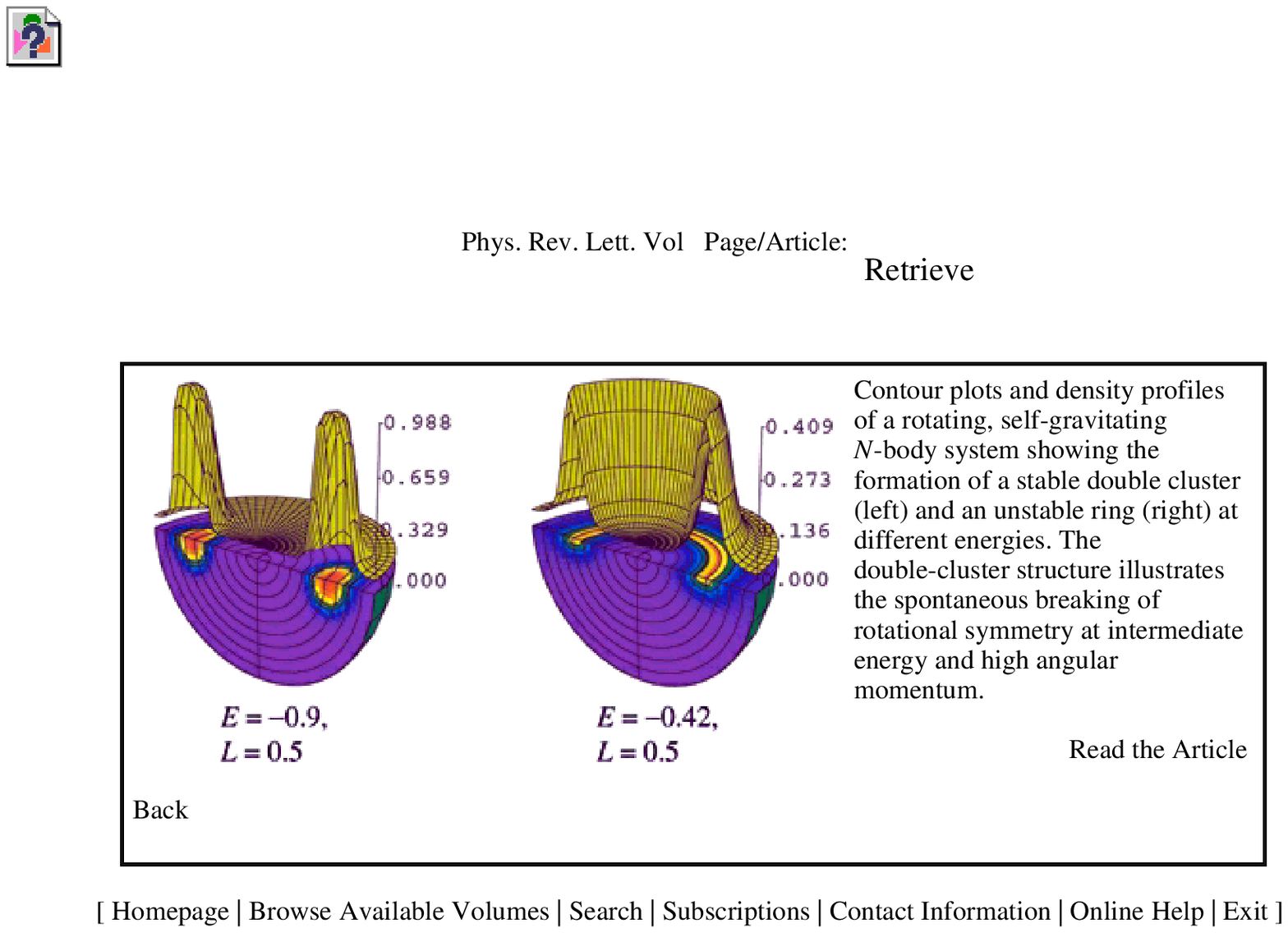}
    \caption{{\bf Phases and Phase-Separation
    in Rotating, Self-Gravitating Systems},
    Physical Review Letters--July 15, 2002, cover-page, by
    (Votyakov, Hidmi, De Martino, Gross)\label{PRL}}
    \end{figure}

\begin{figure}
\includegraphics[bb =72 54 533 690,width=8cm,angle=-90,clip=true]{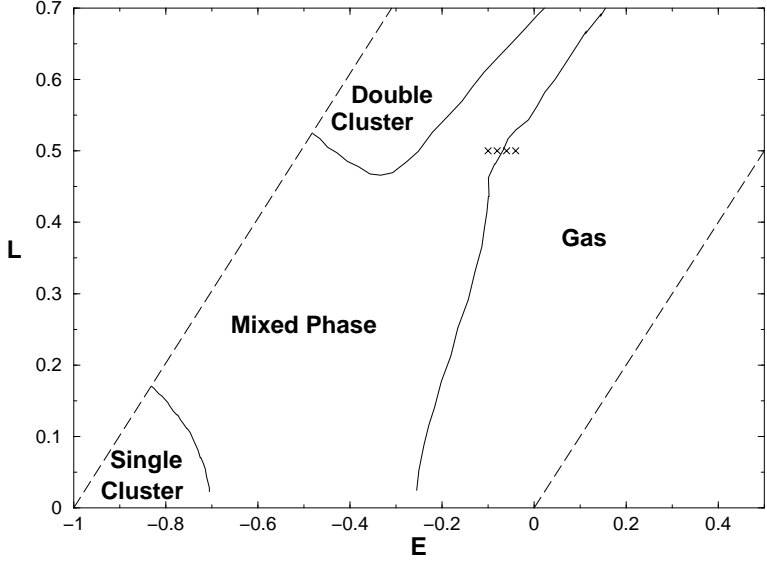}
\caption{{\bf Microcanonical global phase-diagram of a cloud of self-gravitating
and rotating system in a spherical container as function of energy and
angular-momentum.} In this mean-field calculation local densities higher than
the density where nuclear hydrogen burning starts are excluded. Thus,
the gravitational collapse to only {\em visible} stars is followed.
Outside the dashed boundaries only some singular points were calculated
(e.g. see also previous figure). In the mixed phase the largest curvature
$\lambda_1$ of $S(E,L)$ is positive. Consequently the heat
capacity or the correspondent susceptibility is negative. This is
of course well known in astrophysics. However, the new and
important point of our finding is that within microcanonical
thermodynamics this is {\em a generic property of all phase
transitions of first order}, independently of whether there is a short- or a
long-range force that organizes the system. The importance of using the
microcanonical ensemble with angular momentum as the second control
parameter cannot be overemphasized:
The kinetic, "chaotic energy" $E_{kin}=E -E_{pot} - L^2/2I$ is maximized at
large moment of inertia $I$.
Similar equilibrium calculations are done in astro-physics with angular
velocity $\Omega=L/I$ as control parameter ($\sim$ canonical ensemble on the
rotating disk),
whence $E_{kin}=E -E_{pot} - I\times\Omega^2/2$. Then of course, the maximum of
the entropy is at small moment of inertia and only mono-stars are found.
\label{phasediagram}}
\end{figure}

\vspace*{-0.5cm}
    Self-gravitation leads to a non-extensive potential energy $\propto
N^2$.  No thermodynamic limit exists for $E/N$ and no canonical
treatment makes sense. At negative total energies often these systems
have a negative heat capacity.  This was for a long time
considered as an absurd situation within canonical statistical
mechanics with its thermodynamic ``limit''. However, within our
geometric theory this is just a simple example of the
pseudo-Riemannian topology of the microcanonical entropy $S(E,N)$
provided that high densities with their non-gravitational physics,
like nuclear hydrogen burning, are excluded. We treated the
various phases of a self-gravitating cloud of particles as
function of the total energy and angular momentum, c.f.
fig.\ref{PRL} and the quoted PRL-paper.
Clearly these are the most important constraint in
astrophysics. In fig.\ref{phasediagram} the global phase diagram of a rotating,
selfgravitating hydrogen cloud is given.

%%%\clearpage
\section{Conclusion}
\vspace*{-0.5cm}
Entropy has a simple and elementary definition by the {\em area}
$e^{S(E,N,\cdots)}$ of the microcanonical ensemble in the $6N$
dim. phase space. Canonical ensembles are not equivalent to the
micro-ensemble in the most interesting situations:
\begin{enumerate}
\item At phase-separation (\lora heat engines !), one gets
 inhomogeneities, and a negative heat capacity or some other
 negative susceptibility, consequently:
\item  Heat can flow from cold to hot.
\item Phase transitions can be localized sharply and unambiguously in
small classical or quantum systems, there is no need for finite
size scaling to identify the transition.
\item Also really large self-gravitating systems can be addressed.
\end{enumerate}
Entropy rises during the approach to equilibrium, $\Delta S\ge 0$,
also for small mixing systems. i.e. the Second Law is valid even
if the Poincar\'{e} recurrence time is not astronomically large
\cite{gross183,gross192,gross174,gross189}.

With this geometric foundation thermo-statistics applies not only
to extensive systems but also to non-extensive ones which have no
thermodynamic limit. More details are discussed in the references,
see also my WEB-page http://www.hmi.de/people/gross/.

\section{Acknowledgement}
\vspace*{-0.5cm}
The author is grateful to A.Dellafiore, F.Matera, M. Pettini, and
the members of the physics department of Florence,  but especially to
S.Ruffo, for many helpful discussions and their warm hospitality.
He also acknowledges the financial support by the INFN and the
University of Florence.
%%%\newpage

%%%}\end{center}
%%%\end{document}\subsection{Heat can flow from cold to hot}

%%%Bibliographic references with the natbib package:
% Parenthetical: \citep{Bai92} produces (Bailyn 1992).
% Textual: \citet{Bai95} produces Bailyn et al. (1995).
% An affix and part of a reference:
%   \citep[e.g.][Ch. 2]{Bar76}
%   produces (e.g. Barnes et al. 1976, Ch. 2).

%%%\begin{thebibliography}{}

% \bibitem[Names(Year)]{label} or \bibitem[Names(Year)Long names]{label}.
% (\harvarditem{Name}{Year}{label} is also supported.)

%%%\bibitem[]{}

%%%\end{thebibliography}

\end{document}